\documentclass[a4paper,onecolumn,11pt,unpublished]{quantumarticle}
\pdfoutput=1
\usepackage[utf8]{inputenc}
\usepackage[english]{babel}
\usepackage[T1]{fontenc}
\usepackage{amsmath}
\usepackage{hyperref}
\usepackage{braket}
\usepackage[frozencache]{minted}
\usepackage{subcaption}
\usepackage{booktabs}
\usepackage{float}
\usepackage{multirow}
\usepackage{algorithm}
\usepackage{nicematrix}
\usepackage{algpseudocodex}
\usepackage{qcircuit}
\usepackage[numbers,sort&compress]{natbib}

\begin{document}

\title{Automated Auxiliary Qubit Allocation in High-Level Quantum Programming}

\author{Evandro C. R. Rosa}
\orcid{0000-0002-8197-9454}
\email{evandro.crr@posgrad.ufsc.br}

\author{Jerusa Marchi}
\orcid{0000-0002-4864-3764}

\author{Eduardo I. Duzzioni}
\orcid{0000-0002-8971-2033}

\author{Rafael de Santiago}
\orcid{0000-0001-7033-125X}

\affiliation{Universidade Federal de Santa Catarina, Grupo de Computação Quântica, 88040-900 Florianópolis, SC, Brazil}

\begin{abstract}

  We present a method for optimizing quantum circuit compilation by automating the allocation of auxiliary qubits for multi-qubit gate decompositions. This approach is implemented and evaluated within the high-level quantum programming platform Ket. Our results indicate that the decomposition of multi-qubit gates is more effectively handled by the compiler, which has access to all circuit parameters, rather than through a quantum programming API.
  To evaluate the approach, we compared our implementation against Qiskit, a widely used quantum programming platform, by analyzing two quantum algorithms. Using a 16-qubit QPU, we observed a reduction of 87\% in the number of CNOT gates in Grover's algorithm for 9 qubits. For a state preparation algorithm  with 7 qubits, the number of CNOT gates was reduced from $2.8\times10^7$ to $5.7\times10^3$, leveraging additional Ket optimizations for high-level quantum program constructions. Overall, a quadratic reduction in the number of CNOT gates in the final circuit was observed, with greater improvements achieved when more auxiliary qubits were available.
  These findings underscore the importance of automatic resource management, such as auxiliary qubit allocation, in optimizing quantum applications and improving their suitability for near-term quantum hardware.

\end{abstract}

\section{Introduction}

High-level programming languages abstract hardware details from the programmer, allowing them to focus primarily on solution development rather than hardware-specific intricacies~\cite{scottProgrammingLanguagePragmatics2016}. This abstraction has greatly facilitated the creation of computer software and computational solutions, as it shifted the entry point for developers from assembly-level programming to languages closer to the problem domain. A similar paradigm shift, previously observed in classical computing, is now underway in quantum computing. High-level quantum programming languages~\cite{svoreEnablingScalableQuantum2018,siraichiQubitAllocation2018} and platforms~\cite{darosaKetQuantumProgramming2022} aim to abstract quantum hardware, enabling programmers to concentrate more on quantum algorithms design. 

Despite these advancements, high-level quantum programming languages generate instructions that cannot be directly executed on quantum hardware. Similarly to classical computing where processors (CPUs) cannot directly interpret high-level instructions and
source code must be compiled into binary code executable by the CPU~\cite{ahoCompilersPrinciplesTechniques2007}, quantum programs must undergo a compilation process to transform high-level quantum instructions into a sequence of control pulses suitable for execution on a quantum processing unit (QPU).


The compilation process of high-level quantum code can be divided into three stages: decomposition, circuit mapping, and pulse generation, as detailed as follows.
%
In the first stage, gate decomposition~\cite{dasilvaLineardepthQuantumCircuits2022,itenIntroductionUniversalQCompiler2019a,rosaOptimizingGateDecomposition2024,valeCircuitDecompositionMulticontrolled2024,barencoElementaryGatesQuantum1995,itenQuantumCircuitsIsometries2016}, multi-qubit instructions are broken down into one- and two-qubit gates. This step is essential once
quantum computers are limited to a finite set of native gates, whereas high-level code can express operations using an infinite set of gates.
Following,
the circuit must be mapped to the coupling graph of the quantum computer~\cite{chowdhuryQubitAllocationStrategies2024,liTacklingQubitMapping2019,niuHardwareAwareHeuristicQubit2020,willeMQTQMAPEfficient2023,zhuDynamicLookAheadHeuristic2020,zhuVariationAwareQuantumCircuit2023}. Many quantum computing technologies, such as superconducting qubits, do not allow arbitrary interactions between qubits. Therefore, the circuit must be adapted to the hardware's connectivity constraints.
Finally, a pulse sequence is generated~\cite{huangCalibratingSinglequbitGates2023,stefanazziQICKQuantumInstrumentation2022,alexanderQiskitPulseProgramming2020} to be sent to the quantum processor. This step involves translating each gate in the circuit according to a calibration table specific to the quantum hardware. Once the pulse sequence is prepared, the quantum processing unit (QPU) can execute it.

In classical computing, high-level code is designed to be hardware-agnostic, while the compilation process is tailored to the specific requirements of the target processor. This division of responsibilities allows programmers to focus on algorithm design without being concerned with hardware-specific details, as these are managed by the compiler. This approach simplifies software development and results in reusable, maintainable code and enables the compiler to apply a wide range of optimization strategies for improved performance. In this paper, we show that these same principles apply to quantum programming, where similar benefits can be realized through the abstraction of quantum hardware and the use of compiler optimizations.

The decomposition process, being one of the initial steps in preparing high-level quantum code for execution, significantly influences the outcome. While there are countless multi-qubit gates, in high-level quantum programming, we focus on a specific subset: multi-control single-target gates. As discussed in Section~\ref{sec:hlp}, these gates are fundamental building blocks for many quantum algorithms and are among the most commonly used multi-qubit gates. Moreover, constructing multi-qubit gates in terms of controlled single-target gates is a standard practice in quantum programming. For instance, Grover's Diffusion operator, which is part of Grover's algorithm~\cite{groverFastQuantumMechanical1996}, involves a multi-controlled Z gate, and the Quantum Fourier Transform required for Shor's algorithm~\cite{gidneyHowFactor20482021, shorPolynomialTimeAlgorithmsPrime1997} is implemented using controlled Phase gates.

Several algorithms are available for decomposing multi-control single-target quantum gates~\cite{dasilvaLineardepthQuantumCircuits2022, rosaOptimizingGateDecomposition2024, valeCircuitDecompositionMulticontrolled2024, barencoElementaryGatesQuantum1995, itenQuantumCircuitsIsometries2016}, and their efficiency is typically evaluated based on the number of CNOT gates generated during the decomposition. A key consideration in many decompositions is the need for additional qubits, referred to as auxiliary qubits, to facilitate the operation. There is often a trade-off between the number of auxiliary qubits and the number of CNOT gates required, which can be viewed as a balance between space (qubits) and time (CNOTs). Unlike classical computers, which execute multiple unrelated processes in parallel, using additional qubits (space) is not inherently problematic for quantum execution. However, the number of CNOT gates plays a direct role in quantum execution time, which, in turn, impacts decoherence~\cite{preskillQuantumComputingNISQ2018}.

In many quantum programming platforms, the selection of the number of auxiliary qubits used in the decomposition is left to the programmer~\cite{steigerProjectQOpenSource2018, svoreEnablingScalableQuantum2018, bergholmPennyLaneAutomaticDifferentiation2018a, javadi-abhariQuantumComputingQiskit2024}, which may force them to break the application abstraction to enforce a more efficient decomposition. For instance, consider a function that implements a multi-qubit gate in a quantum programming library. If this function uses a multi-controlled gate, the library must know the number of available qubits, which qubits are in use, and whether they are in a clean state to select the most efficient decomposition. If this level of complexity is not enough for the programmer to handle, consider the added difficulty when this multi-controlled gate is itself controlled, as might occur in a high-level programming platform. In this paper, we argue that the selection of auxiliary qubits can be more effectively managed by the compiler, and it is sufficient for the programmer to know which controlled operations have more efficient decompositions without needing to select them manually.

By shifting the responsibility for selecting auxiliary qubits and the decomposition algorithm to the compiler, whenever a multi-controlled gate is encountered, the compiler can determine the number of qubits available on the QPU and how many of those are currently in use. With this information, the compiler can select the most efficient decomposition, considering how many qubits are available to serve as auxiliaries and whether those qubits are in a clean state, as these factors significantly influence the number of CNOT gates required for the decomposition. Once the decomposition algorithm is chosen, the compiler can assign auxiliary qubits based on the interactions between qubits, aiming to minimize the introduction of new interactions that could complicate the subsequent step in the compilation process: circuit mapping.

This paper presents the following key contributions:
\begin{itemize}
  \item An algorithm that autonomously selects the most efficient decomposition for a multi-qubit gate and automatically allocate the auxiliary qubits needed based on the current state of the quantum circuit and QPU.
  \item The implementation of our proposal in the Ket quantum programming platform~\cite{darosaKetQuantumProgramming2022}, demonstrating that the compiler can optimally select the decomposition algorithms to compile a high-level quantum code for execution on a target QPU.
\end{itemize}

The remainder of this paper is structured as follows: Section~\ref{sec:hlp} introduces the high-level quantum programming platform Ket, where our proposal is implemented, and discusses key design decisions that enable the compiler to autonomously select auxiliary qubits for gate decomposition. Section~\ref{sec:decom} presents the decomposition algorithms that allow for the decomposition of any multi-qubit gate written in Ket, along with their time complexity and the requirements in terms of auxiliary qubits. In Section~\ref{sec:aux}, we describe the main contribution of this paper: the algorithm for the automatic selection of auxiliary qubits for multi-controlled gate decomposition. Section~\ref{sec:perf} evaluates our proposal using two quantum algorithms, demonstrating how the automatic selection of auxiliary qubits can reduce the quantum execution time of a program. Finally, Section~\ref{sec:conclusion} presents our conclusions and suggests directions for future work.

\section{High-Level Quantum Programming}
\label{sec:hlp}

Ket~\cite{darosaKetQuantumProgramming2022} is a quantum programming platform that features a Python-embedded high-level programming language named \emph{Ket}, a runtime library written in Rust called \emph{Libket}, and a multithreaded quantum computer simulator known as the \emph{Ket Bitwise Simulator} (KBW). Ket brings the simplicity of Python to quantum programming by building on concepts familiar to classical programming and adding only the necessary functions and types to enable quantum programming. Libket is responsible for handling qubit allocation, quantum circuit construction, and quantum compilation. It is written in Rust for performance and memory safety, which complements Python's ease of use by enhancing the platform's performance.

The execution time of an application written using Ket can be divided into \emph{classical runtime} and \emph{quantum runtime}. The classical runtime begins when the application starts. During this phase, the quantum circuit is defined based on user input (\emph{e.g.}, in Shor's factorization algorithm, the input is the number to be factored), but the actual quantum execution does not occur yet. Each quantum instruction, such as qubit allocation and gate application, triggers a call to Libket, which internally constructs the quantum circuit. Once the classical computer requests the result of a quantum execution, such as the value of a measurement or an expected value, Libket prepares the quantum circuit for execution and sends it to the quantum execution target, which could be a quantum computer or a simulator. This initiates the quantum runtime, which continues until the quantum processor returns the results. By default, Ket uses the KBW simulator as the execution target, unless configured otherwise.

This section continues in Subsection~\ref{subsec:ket}, where we present a brief overview of quantum programming with Ket, illustrating some of the features available for high-level quantum programming. While in Subsection~\ref{subsec:libket}, we discuss how Libket handles controlled gates and how any quantum computation can be represented using a small set of single-qubit gates. For more detailed information, we refer you to the project website\footnote{\url{https://quantumket.org}} and the original paper proposing the platform~\cite{darosaKetQuantumProgramming2022}. Please note that the platform has evolved since the original publication.

\subsection{Ket Programming}
\label{subsec:ket}

A key concept in Ket is that the qubit is a \emph{first-class citizen} of quantum programming. In programming languages, a first-class citizen is a type that supports most operations available to other entities. In Ket, this means that every quantum operation is performed on the qubit. In contrast with Qiskit~\cite{javadi-abhariQuantumComputingQiskit2024}, where operations are appended to a quantum circuit and the circuit is the primary quantum programming object, in Ket, quantum gates and measurements are performed directly on the qubit, eliminating the need to pass the quantum circuit around the program.

As long as a Python function does not measure or allocate a qubit, it can be considered a quantum gate in Ket. This approach facilitates the development of quantum libraries since there is no need for special constructs, such as inheriting a class, to implement a new quantum gate. Moreover, since any quantum gate is reversible, Ket takes advantage of this by offering ways to call a function in reverse, effectively invoking the inverse of a gate. Additionally, since the controlled version of a quantum gate is also a gate, Ket allows calling any function with control qubits. This is a key aspect of high-level quantum programming, as it enables code modularization and reuse.

Figure~\ref{fig:ket_ex} illustrates two quantum programs written using Ket. The code in Figure~\ref{subfig:prepare} implements an arbitrary quantum state preparation, while Figure~\ref{subfig:qft} implements a Quantum Fourier Transform (QFT), a quantum subroutine used in many quantum algorithms, such as Shor's algorithm~\cite{shorPolynomialTimeAlgorithmsPrime1997}, phase estimation~\cite{kitaevQuantumMeasurementsAbelian1995}, solving systems of linear equations~\cite{harrowQuantumAlgorithmLinear2009}, and differential equations~\cite{huQuantumCircuitsPartial2024}. Note that in both codes, the function calls itself recursively, a construct commonly seen in classical programming~\cite{cormenIntroductionAlgorithms2009} but less frequently in quantum programming. In Figure~\ref{subfig:prepare}, there are no explicit-controlled gates being called, whereas in Figure~\ref{subfig:qft}, there is a Phase gate call inside a \texttt{with control} statement, resulting in a controlled-Phase gate. Although, in Figure~\ref{subfig:prepare}, the recursive call is inside a \texttt{with control} statement; since the function is a quantum gate, it will call the controlled version of itself. In Figure~\ref{subfig:qft}, as any quantum gate is reversible, the implementation of a QFT can automatically derive the Inverse QFT using Ket's \texttt{adj()} function.

Controlled quantum gates arise naturally in high-level quantum programming, as illustrated in Figure~\ref{subfig:prepare}. Considering how Libket implements controlled operations, calling the gate in Figure~\ref{subfig:prepare} will result in a multi-controlled $R_Y$ gate and a multi-controlled Phase gate with an increasing number of control qubits. The Pauli~$X$ gates inside the \texttt{with around} statement do not result in multi-controlled NOT gates due to Ket's optimization~\cite{rosaOptimizingGateDecomposition2024}.

\begin{figure}[htbp]
  \centering
  \begin{subfigure}[c]{0.495\textwidth}
    \centering
    \begin{minipage}{.98\linewidth}
      \begin{minted}[frame=lines,fontsize=\scriptsize,breaklines]{py}
def prepare(
    qubits: Quant,
    prob: ParamTree | list[float],
    amp: list[float] | None = None,
):
    if not isinstance(prob, ParamTree):
        prob = ParamTree(prob, amp)

    head, *tail = qubits
    RY(prob.value, head)
    if prob.is_leaf():
        with around(X, head):
            PHASE(prob.phase0, head)
        PHASE(prob.phase1, head)
        return

    with around(X, head):
        with control(head):
            prepare(tail, prob.left)
    with control(head):
        prepare(tail, prob.right)
      \end{minted}
    \end{minipage}
    \caption{Arbitrary quantum state preparation. For the implementation of class \texttt{ParamTree}, see Figure~\ref{subfig:prepara:ket}.}
    \label{subfig:prepare}
  \end{subfigure}
  \begin{subfigure}[c]{0.495\textwidth}
    \centering
    \begin{minipage}{.98\linewidth}
      \begin{minted}[frame=lines,fontsize=\scriptsize,breaklines]{py}
def qft(qubits: Quant, do_swap: bool = True):
    if len(qubits) == 1:
        H(qubits)
    else:
        *init, last = qubits
        ket.H(last)

        for i, c in enumerate(reversed(init)):
            with control(c):
                PHASE(pi / 2 ** (i + 1), last)

        qft(init, do_swap=False)

    if do_swap:
        for i in range(len(qubits) // 2):
            SWAP(
                qubits[i],
                qubits[size - i - 1],
            )

iqft = adj(qft)
      \end{minted}
    \end{minipage}
    \caption{Quantum Fourier Transformation (QFT) with automatically Inverse QFT derivation.}
    \label{subfig:qft}
  \end{subfigure}

  \caption{Example of code implementation using the high-level quantum programming platform Ket.}
  \label{fig:ket_ex}
\end{figure}

\subsection{Libket Runtime Library}
\label{subsec:libket}

Libket provides a limited set of single-qubit gates, including the Pauli gates, rotation gates, the Phase gate, and the Hadamard gate. This set is sufficient to perform any quantum operation on a single qubit. To achieve universal quantum computation, Libket allows any gate to be called with control qubits. While the ability to call a Pauli~$X$ gate with a single control qubit enables universal quantum computation, Libket also allows an arbitrary number of control qubits to be added to any of the available quantum gates, facilitating high-level quantum programming.

Using rotation gates, any single-qubit operation can be implemented, specifically any special unitary operation ($SU(2)$). The $SU(2)$ gates are a subset of single-qubit gates ($U(2)$). For any single-qubit gate, there exists a $SU(2)$ gate that implements the same operation, ignoring the global phase. However, while the global phase can usually be disregarded in normal gate applications, it becomes a relative phase when the gate is called with controls. For example, consider the $\sqrt{X}$ gate, which is equivalent to a $R_X(\tfrac{\pi}{2})$ gate with a global phase of $e^{i\tfrac{\pi}{4}}$. When calling a controlled version of this gate, the absence of a global phase leads to the following inequality:
\begin{equation}
  C\sqrt{X} = \frac{1}{2}
  \begin{bNiceMatrix}
    2 & 0 & 0   & 0   \\
    0 & 2 & 0   & 0   \\
    0 & 0 & 1+i & 1-i \\
    0 & 0 & 1-i & 1+i
  \end{bNiceMatrix} \neq
  CR_X(\tfrac{\pi}{2}) =
  \frac{1}{\sqrt{2}}
  \begin{bNiceMatrix}
    \sqrt{2} & 0        & 0  & 0  \\
    0        & \sqrt{2} & 0  & 0  \\
    0        & 0        & 1  & -i \\
    0        & 0        & -i & 1
  \end{bNiceMatrix}.
\end{equation}

While the ability to add control qubits to the single-qubit gates in Libket is sufficient for universal quantum computation, the ability to easily describe any single-qubit unitary operation can greatly facilitate high-level quantum programming. To enable the implementation of quantum gates like $\sqrt{X}$, Libket allows the addition of a global phase to any quantum gate. In Ket, this functionality is provided through the \texttt{global\_phase()} function, which can be used as a function decorator. Implementing global phase does not require adding new gates to Libket. For example, the $\sqrt{X}$ gate can be implemented in Ket using the code in Figure~\ref{fig:sx}. When the gate is called normally, only the $R_X(\tfrac{\pi}{2})$ is applied to the qubits. However, when the gate is called with control qubits, a multi-controlled Phase gate is applied to the control qubits to correct the phase. Global phase can also be added to multi-qubit gates. For instance, if the gate in Figure~\ref{fig:sx} is called with two control qubits, it will result in the following gates in Libket:
\begin{equation}
  C^2\sqrt{X} = (CP(\tfrac{\pi}{4})\otimes I) \cdot C^2R_X(\tfrac{\pi}{2}).
\end{equation}

\begin{figure}[htbp]
  \centering
  \begin{minipage}{.46\linewidth}
    \begin{minted}[frame=lines,breaklines]{py}
@global_phase(pi / 4)
def sqrt_x(qubit):
    return RX(pi / 2, qubit)
    \end{minted}
  \end{minipage}
  \caption{$\sqrt{X}$ gate implementation in Ket.}
  \label{fig:sx}
\end{figure}

Ket allows the implementation of any quantum operation up to a global phase, and every controlled operation in Ket results in a controlled version of the gates implemented in Libket. As a result, the compiler has fewer decompositions to manage. While it is possible to call a controlled version of operations like the Quantum Fourier Transform or other quantum gates, the outcomes will only involve multi-controlled versions of Pauli gates, rotation gates, the Hadamard gate, or the Phase gate. This approach enables Libket to automatically allocate auxiliary qubits, as the decomposition process is handled by the compiler rather than being part of the standard programming API. If the decomposition were handled by the programming API--where it is explicitly coded in Ket and Libket merely places the gates--the compiler would lack the necessary information to allocate auxiliary qubits automatically. Furthermore, if the programmer were responsible for the decomposition, they would need to implement mechanisms to select the appropriate decomposition and manage varying numbers of auxiliary qubits while ensuring that the qubits are in the correct state.

In the next section, we present the decomposition algorithms implemented in Libket, which enable the decomposition of any multi-qubit gate defined using Ket's high-level quantum constructions. In Section~\ref{sec:aux}, we discuss how Libket selects the appropriate decomposition based on the execution target and the state of the quantum circuit, aiming to minimize the number of CNOT gates in the final compiled circuit.

\section{Gate Decomposition Algorithms}
\label{sec:decom}

In this section, we present the decomposition algorithms implemented in Libket that enable the decomposition of any multi-qubit quantum gate. Although this is not the primary focus of the paper, we aim to present the state-of-the-art in quantum gate decomposition, as the performance of these algorithms directly impacts the classical CPU time required for compiling high-level quantum code as well as the quantum execution time. Table~\ref{tab:decom} summarizes the algorithms implemented in Libket for decomposing an arbitrary $C^nU$ gate, where the $U$ gate is classified as a Pauli gate, rotation gate, Hadamard gate, or Phase gate.

\begin{table}[htpb]
  \caption{Decomposition algorithms implemented in Libket, sorted by time complexity.}
  \label{tab:decom}
  \centering
  \scriptsize
  \begin{tabular}{clccc}
    \toprule
    \multirow{2}{*}{Quantum Gate}   & \multicolumn{4}{c}{Decomposition Algorithm}                                       \\
    \cmidrule(r){2-5}
    & \multicolumn{1}{c}{Name}                    & N\# Auxiliary Qubits & Auxiliary Qubit State & N\# CNTOs \\
    \midrule
    \midrule
    \multirow{4}{*}{Pauli Gates}    & \multirow{2}{*}{V-Chain~\cite{barencoElementaryGatesQuantum1995}}                    & $n-2$           & Clean & $2n$      \\
    \cmidrule(lr){3-5}

    &                                             & $n-2$           & Dirty & $4n$      \\
    \cmidrule(r){2-5}

    & \multirow{2}{*}{Single-Aux~\cite{barencoElementaryGatesQuantum1995}}                 & $1$             & Clean & $12n$     \\
    \cmidrule(lr){3-5}
    &                                             & $1$             & Dirty & $16n$     \\
    \cmidrule(r){2-5}
    & Linear-Depth~\cite{dasilvaLineardepthQuantumCircuits2022}                                & $0$             & --    & $n^2$     \\
    \cmidrule(lr){1-5}

    \multirow{2}{*}{Rotation Gates} & Network~\cite[p. 183]{nielsenQuantumComputationQuantum2010}                                     & $n-1$           & Clean & $2n$      \\
    \cmidrule(r){2-5}
    & $SU(2)$~\cite{valeCircuitDecompositionMulticontrolled2024}                                     & $0$             & --    & $16n$     \\
    \cmidrule(lr){1-5}
    \multirow{3}{*}{Phase Gate}     & Network~\cite[p. 183]{nielsenQuantumComputationQuantum2010}                                     & $n-1$           & Clean & $2n$      \\
    \cmidrule(r){2-5}
    & $SU(2)$ Rewrite~\cite{rosaOptimizingGateDecomposition2024}                             & $1$             & Clean & $16n$     \\
    \cmidrule(r){2-5}
    & Linear-Depth~\cite{dasilvaLineardepthQuantumCircuits2022}                                   & $0$             & --    & $n^2$     \\
    \cmidrule(lr){1-5}
    \multirow{3}{*}{Hadamard Gate}  & Network~\cite[p. 183]{nielsenQuantumComputationQuantum2010}                                     & $n-1$           & Clean & $2n$      \\
    \cmidrule(r){2-5}
    & $SU(2)$ Rewrite~\cite{rosaOptimizingGateDecomposition2024}                               & $1$             & Clean & $32n$     \\
    \cmidrule(r){2-5}
    & Linear-Depth~\cite{dasilvaLineardepthQuantumCircuits2022}                                   & $0$             & --    & $n^2$     \\
    \bottomrule
  \end{tabular}
\end{table}

\paragraph{Specific Decompositions}

In certain cases, decomposition is possible without auxiliary qubits. For example, Pauli gates with 2 or 3 control qubits can be optimally decomposed into 6 and 14 CNOT gates, respectively. Additionally, for an arbitrary single-qubit gate with one control qubit, the $CU(2)$ decomposition, illustrated in Figure~\ref{fig:cu2}, can be used.

\begin{figure}[htpb]
  \begin{equation*}
    \Qcircuit @C=1em @R=.8em {
      &  \ctrl{1} & \qw & = & & \qw      & \ctrl{1} & \qw      & \ctrl{1} & \gate{P(\theta)} & \qw \\
      &  \gate{U} & \qw &   & & \gate{C} & \targ    & \gate{B} & \targ    & \gate{A}         & \qw
    }
  \end{equation*}
  \caption{$CU(2)$ decomposition, where $U = e^{i\theta}AXBXC$~\cite[Corollary 4.2]{nielsenQuantumComputationQuantum2010}.}
  \label{fig:cu2}
\end{figure}

\paragraph{Network Decomposition}

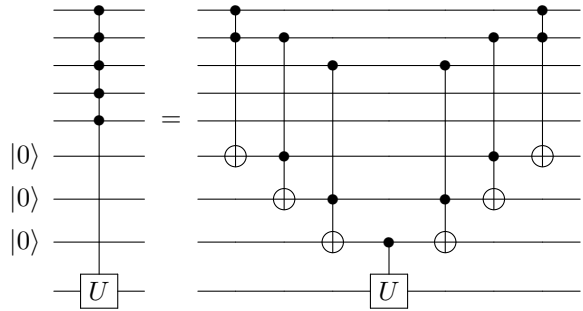
\begin{figure}[htpb]
  \small
  \begin{equation*}
    \Qcircuit @C=1em @R=.8em {
      &  \ctrl{1} & \qw & &  &  \ctrl{1} & \qw      & \qw       & \qw      & \qw      & \qw      & \ctrl{1} & \qw \\
      &  \ctrl{1} & \qw & &  &  \ctrl{4} & \ctrl{4} & \qw       & \qw      & \qw      & \ctrl{4} & \ctrl{4} & \qw \\
      &  \ctrl{1} & \qw & &  &  \qw      & \qw      & \ctrl{4}  & \qw      & \ctrl{4} & \qw      & \qw      & \qw \\
      &  \ctrl{1} & \qw & &  &  \qw      & \qw      & \qw       & \qw      & \qw      & \qw      & \qw      & \qw \\
      &  \ctrl{4} & \qw &=&  &  \qw      & \qw      & \qw       & \qw      & \qw      & \qw      & \qw      & \qw \\
      \lstick{\ket{0}}&  \qw      & \qw & &  &  \targ    & \ctrl{1} & \qw       & \qw      & \qw      & \ctrl{1} & \targ    & \qw \\
      \lstick{\ket{0}}&  \qw      & \qw & &  &  \qw      & \targ    & \ctrl{1}  & \qw      & \ctrl{1} & \targ    & \qw      & \qw \\
      \lstick{\ket{0}}&  \qw      & \qw & &  &  \qw      & \qw      & \targ     & \ctrl{1} & \targ    & \qw      & \qw      & \qw \\
      &  \gate{U} & \qw & &  &  \qw      & \qw      & \qw       & \gate{U} & \qw      & \qw      & \qw      & \qw
    }
  \end{equation*}
  \caption{Network decomposition. The gate in the middle of the network can be further decomposed as shown in Figure~\ref{fig:cu2}.}
  \label{fig:network}
\end{figure}

The network decomposition~\cite[p. 183]{nielsenQuantumComputationQuantum2010} can be applied to any multi-controlled, single-target quantum gate. In Libket, this decomposition is particularly used for rotation gates, Phase gates, and the Hadamard gate, utilizing $n-1$ clean auxiliary qubits for a gate with $n$ control qubits. This decomposition, depicted in Figure~\ref{fig:network}, results in at most $2n$ CNOT gates. For Toffoli gates ($C^2X$), the approximate Toffoli decomposition~\cite[Lemma 9]{itenQuantumCircuitsIsometries2016} using 3 CNOTs is used instead of the exact Toffoli decomposition that requires 6 CNOTs.

\paragraph{V-Chain Decomposition}

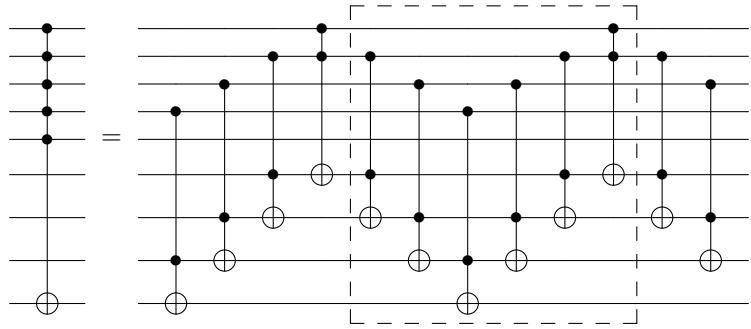
\begin{figure}[htpb]
  \small
  \begin{equation*}
    \Qcircuit @C=1em @R=.8em {
      &  \ctrl{1} & \qw & & & \qw      & \qw      & \qw      & \ctrl{1} & \qw      & \qw       & \qw      & \qw      & \qw      & \ctrl{1} & \qw      & \qw       & \qw \\
      &  \ctrl{1} & \qw & & & \qw      & \qw      & \ctrl{4} & \ctrl{4} & \ctrl{4} & \qw       & \qw      & \qw      & \ctrl{4} & \ctrl{4} & \ctrl{4} & \qw       & \qw \\
      &  \ctrl{1} & \qw & & & \qw      & \ctrl{4} & \qw      & \qw      & \qw      & \ctrl{4}  & \qw      & \ctrl{4} & \qw      & \qw      & \qw      & \ctrl{4}  & \qw \\
      &  \ctrl{1} & \qw & & & \ctrl{4} & \qw      & \qw      & \qw      & \qw      & \qw       & \ctrl{4} & \qw      & \qw      & \qw      & \qw      & \qw       & \qw \\
      &  \ctrl{4} & \qw &=& & \qw      & \qw      & \qw      & \qw      & \qw      & \qw       & \qw      & \qw      & \qw      & \qw      & \qw      & \qw       & \qw \\
      &  \qw      & \qw & & & \qw      & \qw      & \ctrl{1} & \targ    & \ctrl{1} & \qw       & \qw      & \qw      & \ctrl{1} & \targ    & \ctrl{1} & \qw       & \qw \\
      &  \qw      & \qw & & & \qw      & \ctrl{1} & \targ    & \qw      & \targ    & \ctrl{1}  & \qw      & \ctrl{1} & \targ    & \qw      & \targ    & \ctrl{1}  & \qw \\
      &  \qw      & \qw & & & \ctrl{1} & \targ    & \qw      & \qw      & \qw      & \targ     & \ctrl{1} & \targ    & \qw      & \qw      & \qw      & \targ     & \qw \\
      &  \targ    & \qw & & & \targ    & \qw      & \qw      & \qw      & \qw      & \qw       & \targ    & \qw      & \qw      & \qw      & \qw      & \qw       & \qw
      \gategroup{1}{10}{9}{15}{1.5em}{--}
    }
  \end{equation*}
  \caption{V-Chain decomposition. Gates inside the dashed box are not required when using clean auxiliary qubits.}
  \label{fig:vchain}
\end{figure}

One of the most efficient decompositions for multi-controlled Pauli~$X$ gates is the V-Chain algorithm~\cite{barencoElementaryGatesQuantum1995}, which requires $n-2$ auxiliary qubits to decompose a gate with $n$ control qubits. The decomposition results in at most $2n$ CNOTs when clean auxiliary qubits are available, and $4n$ CNOTs when dirty auxiliary qubits are used. The reduction in CNOTs with clean auxiliaries occurs because some gates can be skipped when the auxiliary qubits are known to be in the $\ket{0\cdots0}$ state. Figure~\ref{fig:vchain} illustrates the V-Chain decomposition algorithm, where only the gates within the dashed box are required when using clean auxiliaries. Although this decomposition is primarily used for multi-controlled Pauli~$X$ gates, it can be adapted for other Pauli gates by applying a basis change to the target qubit. The approximate Toffoli decomposition~\cite{itenQuantumCircuitsIsometries2016} is used when the gate is not applied to the original target.

\paragraph{Single-Aux Decomposition}

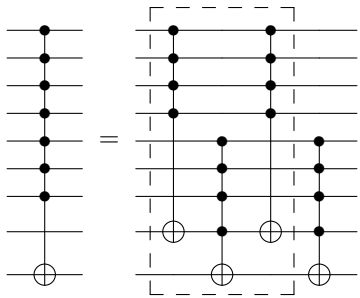
\begin{figure}[htpb]
  \small
  \begin{equation*}
    \Qcircuit @C=1em @R=.8em {
      &  \ctrl{1} & \qw & & & \ctrl{1} & \qw      & \ctrl{1} & \qw      & \qw \\
      &  \ctrl{1} & \qw & & & \ctrl{1} & \qw      & \ctrl{1} & \qw      & \qw \\
      &  \ctrl{1} & \qw & & & \ctrl{1} & \qw      & \ctrl{1} & \qw      & \qw \\
      &  \ctrl{1} & \qw & & & \ctrl{4} & \qw      & \ctrl{4} & \qw      & \qw \\
      &  \ctrl{1} & \qw &=& & \qw      & \ctrl{1} & \qw      & \ctrl{1} & \qw \\
      &  \ctrl{1} & \qw & & & \qw      & \ctrl{1} & \qw      & \ctrl{1} & \qw \\
      &  \ctrl{2} & \qw & & & \qw      & \ctrl{1} & \qw      & \ctrl{1} & \qw \\
      &  \qw      & \qw & & & \targ    & \ctrl{1} & \targ    & \ctrl{1} & \qw \\
      &  \targ    & \qw & & & \qw      & \targ    & \qw      & \targ    & \qw
      \gategroup{1}{6}{9}{8}{1.5em}{--}
    }
  \end{equation*}
  \caption{Single-Aux decomposition. Gates inside the dashed box are not required when using clean auxiliary qubits. The V-Chain decomposition with dirty auxiliary qubits is used to decompose the gates on the right side of the equation.}
  \label{fig:single-Aux}
\end{figure}

Using a single auxiliary qubit, any Pauli gate with $n$ control qubits can be decomposed into at most $16n$ CNOT gates with a dirty auxiliary~\cite{barencoElementaryGatesQuantum1995} or $12n$ with a clean auxiliary~\cite{rosaOptimizingGateDecomposition2024}. As depicted in Figure~\ref{fig:single-Aux}, this decomposition breaks the $n$-controlled Pauli~$X$ gate into three or four $\frac{n}{2}$-controlled Pauli~$X$ gate decompositions. In the case of clean auxiliary qubits, only the gates within the dashed box are necessary. The V-Chain decomposition can be applied to each of the $\frac{n}{2}$-controlled Pauli~$X$ gates.

\paragraph{$SU(2)$ Decomposition}

\begin{figure}
  \begin{equation*}
    \Qcircuit @C=1em @R=1em {
      & \qw {/} & \ctrl{2}            & \qw &   & & \qw {/}^{k_0} & \qw              & \ctrl{2} & \qw      & \qw      & \qw              & \ctrl{2} & \qw      & \qw     & \qw              & \qw              & \qw \\
      & \qw {/} & \ctrl{1}            & \qw & = & & \qw {/}^{k_1} & \qw              & \qw      & \qw      & \ctrl{1} & \qw              & \qw      & \qw      & \ctrl{1}& \qw              & \qw              & \qw \\
      & \qw     & \gate{U}             & \qw &   & & \qw           & \gate{V^\dagger} & \targ    & \gate{A} & \targ    & \gate{A^\dagger} & \targ    & \gate{A} & \targ   & \gate{A^\dagger} & \gate{V} & \qw \\
    }
  \end{equation*}
  \caption{$SU(2)$ decomposition with $n$ control qubits, where $k_0 = \lfloor{\tfrac{n}{2}}\rfloor$, $k_1 = n - k_0$, $U = VDV^\dagger$, and $D = A^\dagger X A X A^\dagger X A X$.}
  \label{fig:su2}
\end{figure}
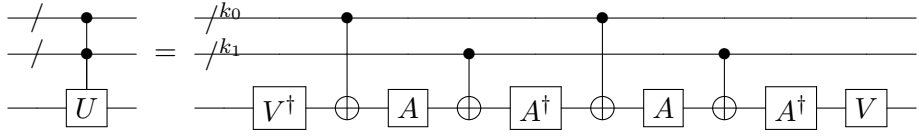

For gates within the $SU(2)$ group, such as rotation gates, with $n$ control qubits, the decomposition shown in Figure~\ref{fig:su2} can be used, resulting in at most $16n$ CNOT gates~\cite{valeCircuitDecompositionMulticontrolled2024}. This decomposition does not require auxiliary qubits and is performed using four V-Chain decompositions.

\paragraph{$SU(2)$ Rewrite Decomposition}

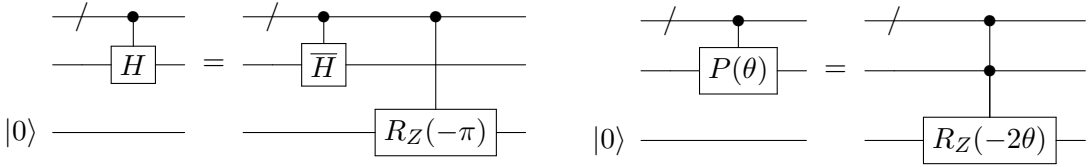
\begin{figure}
  \begin{subfigure}[c]{0.495\textwidth}
    \begin{equation*}
      \Qcircuit @C=1em @R=.8em {
        & {/} \qw & \ctrl{1} & \qw &   & & {/} \qw & \ctrl{1}                                & \ctrl{2}             & \qw \\
        &     \qw & \gate{H} & \qw & = & &     \qw & \gate{\overline{H}}      & \qw                  & \qw \\
        \lstick{\ket{0}} &     \qw      & \qw & \qw &   & &     \qw   & \qw & \gate{R_Z(-\pi)} & \qw
      }
    \end{equation*}
    \caption{Hadamard gate decomposition with $SU(2)$ Rewrite, where $\overline{H} = R_X(\pi)R_Y(\tfrac{\pi}{2})$.}
  \end{subfigure}
  \begin{subfigure}[c]{0.495\textwidth}
    \begin{equation*}
      \Qcircuit @C=1em @R=.8em {
        & {/} \qw & \ctrl{1}         & \qw &   & & {/} \qw & \ctrl{2}             & \qw \\
        &     \qw & \gate{P(\theta)} & \qw & = & &     \qw & \ctrl{1}             & \qw \\
        \lstick{\ket{0}} &     \qw & \qw              & \qw &   & &     \qw & \gate{R_Z(-2\theta)} & \qw \\
      }
    \end{equation*}
    \caption{Phase gate decomposition with $SU(2)$ Rewrite.}
  \end{subfigure}
  \caption{$SU(2)$ Rewrite decompositions.}
  \label{fig:su2r}
\end{figure}

For gates outside the $SU(2)$ group, any single-qubit gate $U$ can be expressed as $U = e^{i\varphi} \overline{U}$, where $\overline{U} \in SU(2)$. This allows us to use the $SU(2)$ decomposition for $\overline{U}$ and apply a multi-controlled $R_Z(-2\varphi)$ decomposition to fix the phase with a single clean auxiliary qubit~\cite{rosaOptimizingGateDecomposition2024}. In Libket, this decomposition is necessary for the Hadamard and Phase gates. Figure~\ref{fig:su2r} illustrates this decomposition algorithm, particularly for the Hadamard and Phase gates. Notably, the Phase gate can be rewritten as a single multi-controlled $R_Z$ gate. For a gate with $n$ control qubits, the decomposition results in at most $32n$ CNOT gates for the Hadamard gate and $16n$ CNOT gates for the Phase gate.

\paragraph{Linear Depth Decomposition}

When no auxiliary qubits are available for decomposing a gate that is not in $SU(2)$, the Linear Depth decomposition can be used~\cite{dasilvaLineardepthQuantumCircuits2022}. This is the fallback decomposition used in worst-case scenarios, ensuring that any multi-qubit gate described in Ket can be decomposed. Although this decomposition results in a quadratic number of CNOT gates relative to the number of control qubits, it produces a linear-depth circuit, meaning it can be executed in linear time by the QPU despite the quadratic CPU runtime during compilation\footnote{Execution time and compilation time differ; for example, execution may scale linearly due to parallel gate operations, while compilation can scale quadratically as gates are arranged and scheduled.}.

\section{Automatic Auxiliary Allocation}
\label{sec:aux}

The decomposition of multi-controlled single-target gates takes place during classical runtime, when Libket can access detailed information about the target QPU and classical inputs. This includes the number of qubits available and details about the final quantum circuit. Such information is crucial, as it identifies the qubits directly involved in the gate operation, and, most importantly, highlights the qubits that are not involved in the quantum execution, which can be utilized as auxiliary qubits. This section details how Libket automatically selects the optimal decomposition algorithm and allocates auxiliary qubits accordingly.

In our approach, we distinguish between \emph{main qubits} and \emph{auxiliary qubits}. Main qubits are those allocated by the programmer from a quantum process and are mapped to physical qubits during the compilation process. In contrast, auxiliary qubits are managed by the runtime library for the decomposition and are inaccessible to the programmer. During compilation, auxiliary qubits are not directly mapped to physical qubits; instead, they are associated with main qubits. This section begins with a discussion of how decomposition algorithms are selected in Subsection~\ref{subsec:selct} and concludes with how auxiliary qubits are allocated to main qubits in Section~\ref{subsec:alloc}. Performance evaluations of this technique are presented in Section~\ref{sec:perf}.

\subsection{Decomposition Selection}
\label{subsec:selct}

As presented in Section~\ref{sec:decom}, some decompositions require auxiliary qubits to be in the $\ket{0}$ state, while others are independent of the qubits' initial state. Auxiliary qubits that start in the $\ket{0}$ state are called clean, while those that do not have a specific initial state are called dirty. It is important to note that dirty auxiliary qubits are not in a mixed state, but rather in a pure state that is unknown. The determination of whether auxiliary qubits are available for decomposition strongly depends on whether the auxiliary qubits are clean or dirty.

To select the best decomposition for a multi-controlled gate, we first sort the possible algorithms by their time complexity--in other words, by how many CNOTs the decomposition requires for a given number of control qubits. As discussed in Section~\ref{sec:decom}, decompositions that require more auxiliary qubits typically require fewer CNOTs to implement, and those that use clean qubits are generally faster than those using dirty qubits.

We begin by checking if there are enough qubits available for the most efficient decomposition. If there are not, we move on to the next option. If no auxiliary qubits are available, we can always fall back to a decomposition with quadratic time complexity in the worst case~\cite{dasilvaLineardepthQuantumCircuits2022}, such as with Pauli gates, the Hadamard gate, and the Phase gate.

The calculation to determine if there are enough qubits available for the decomposition is straightforward for clean qubits. Since the runtime library knows how many qubits have been allocated by the program and which of those have been used, it only needs to determine if the number of unused qubits is greater than or equal to the number of auxiliary qubits needed for the decomposition. If so, there are enough qubits available.

To evaluate if enough dirty qubits are available for decomposition, we consider the qubits the auxiliary qubits will interact with, referred to as the \emph{interaction group}. The interaction group includes the qubits that form the original controlled operation. To proceed with decomposition, the following condition must be met: the total number of auxiliary qubits required, plus the number of qubits in the interaction group, must not exceed the total number of qubits available on the quantum processing unit (QPU). 

Clean qubits can also be used as dirty auxiliary qubits because, at the end of the decomposition process, these qubits are restored to their initial state, effectively remaining ``clean''. Once we confirm that the QPU has sufficient qubits available, including clean qubits repurposed as dirty auxiliaries, the decomposition proceeds using the auxiliary qubits.

Regardless of whether the auxiliary qubits are clean or dirty, they are treated as new qubits. For instance, they appear as new lines in the quantum circuit, and every quantum gate applied to the auxiliary will be applied to this new line. The auxiliary qubits will only be mapped to main qubits once the decomposition is complete, as will be explained in the next subsection. This is necessary because the information about which qubits the auxiliary interacts with is important for the mapping.

Consider the following example to illustrate how the decomposition selection is made. Suppose we have the statement \texttt{ctrl(qubits[:-1], Z)(qubits[-1])}, which represents a multi-controlled Pauli~$Z$ gate, where \texttt{qubits} is a list of $n$ qubits. Additionally, assume that this operation is being executed on a 12-qubit QPU. We will examine several scenarios based on the decompositions listed in Table~\ref{tab:decom}:

\begin{itemize}
  \item  For $n \leq 4$: In this scenario, there are known decompositions that do not require any auxiliary qubits.
  \item  For $4 < n \leq 7$: The V-Chain decomposition can be used, as there are sufficient qubits available to serve as auxiliary qubits.
  \item  For $7 < n \leq 11$: The best option is the Single-Aux decomposition, since the V-Chain method demands more qubits than are available on the QPU.
  \item  For $n = 12$: The Linear-Depth decomposition is the sole option, as all qubits are engaged in the gate operation.
\end{itemize}

To determine whether the decomposition can be performed with clean auxiliary qubits or not, consider the case where $n = 6$. In this scenario, the V-Chain decomposition is applicable. If, in addition to the qubits involved in the operation, the program has allocated at most 3 additional qubits, the decomposition can use clean auxiliary qubits, resulting in a more efficient circuit. This is possible because the operation uses 6 qubits, the program has allocated 3 additional qubits, and the decomposition requires 3 auxiliary qubits, bringing the total to 12 qubits--out of the 12-qubits QPU. However, if the program has allocated between 4 and 6 additional qubits, at least one of the auxiliary qubits will be dirty, resulting in a decomposition with dirty qubits.

\subsection{Qubit Allocation}
\label{subsec:alloc}

Once the multi-controlled gate has been decomposed using auxiliary qubits, the next step is to allocate these auxiliary qubits to main qubits. This allocation takes place following the decomposition, leveraging information about the interactions involving the auxiliary qubits to enable informed decision-making. The goal is to allocate auxiliary qubits in a manner that minimizes the introduction of new interactions between qubits. This strategy aims to reduce the number of SWAP operations required to adapt the quantum circuit to the QPU's coupling graph, which is the subsequent step in the compilation process.

In our approach, a set of auxiliary qubits serves a single purpose: to assist in the decomposition of a multi-controlled quantum gate. We group auxiliary qubits that participate in the same decomposition into what we term an \emph{auxiliary group}. The lifetime of an auxiliary qubit ends after its allocation, and new decompositions require the allocation of new auxiliary qubits. Although the allocation process may reuse the same main qubits for different auxiliary groups, there is no logical correlation between distinct auxiliary groups.

Auxiliary qubits can only be allocated once they have been returned to their original state, whether they are clean or dirty. The decomposition algorithms are designed with this constraint in mind. Additionally, a single main qubit cannot host more than one auxiliary qubit per decomposition, so this must be tracked during the allocation process. It is important to note that decompositions are performed sequentially, ensuring that auxiliary groups do not overlap.

Algorithm~\ref{alg:alloc} illustrates the process of determining which main qubits will host the auxiliary qubits in a given group. The algorithm takes as input the auxiliary qubits group $\mathcal{A}$ and their interaction group $\mathcal{I}$, which is empty if the qubits are in a clean state. The algorithm begins by initializing an empty set $\mathcal{Q}$ (line~\ref{line:alloc:new_set}), which will store the main qubits that already have an auxiliary qubit from $\mathcal{A}$ allocated to them. The algorithm then iterates over each auxiliary qubit $a$ in $\mathcal{A}$ (line~\ref{line:alloc:forall}), using the function \textsc{SelectMain} from Algorithm~\ref{alg:aux} to determine the main qubit $q$ to which the auxiliary qubit $a$ will be allocated  (line~\ref{line:alloc:selct}). After determining the appropriate main qubit $q$, it is added to the set $\mathcal{Q}$ (line~\ref{line:alloc:update}), and all quantum gates associated with $a$ are moved to $q$ (line~\ref{line:alloc:move}). During this process, any two-qubit gates that involve $a$ are updated to instead use $q$. Once this is done, $a$ is discarded. By the end of this process, all auxiliary qubits will have been allocated to main qubits, with the quantum circuit reflecting these allocations. Since the auxiliary qubits are discarded after allocation, only the main qubits remain in the circuit when it is ready for the circuit mapping compilation phase.

\begin{figure}[htpb]
  \centering
  \begin{minipage}{.5\linewidth}
    \begin{algorithm}[H]
      \caption{Allocate auxiliary qubits}
      \label{alg:alloc}
      \begin{algorithmic}[1]
        \Procedure{AllocateAux}{$\mathcal{A}$, $\mathcal{I}$}

        \State $\mathcal{Q} \gets \emptyset$ \label{line:alloc:new_set}

        \ForAll {$a \in \mathcal{A}$} \label{line:alloc:forall}
        \State $q \gets$ \Call{SelectMain}{$a$, $\mathcal{A}$, $\mathcal{I}$, $\mathcal{Q}$} \label{line:alloc:selct}
        \State $\mathcal{Q} \gets \mathcal{Q} \cup \{q\}$ \label{line:alloc:update}
        \State \Call{MoveGatesFromTo}{$a$, $q$} \label{line:alloc:move}
        \EndFor

        \EndProcedure

      \end{algorithmic}
    \end{algorithm}
  \end{minipage}
\end{figure}

Algorithm~\ref{alg:aux} is central to the allocation process, as it determines which main qubit will host a given auxiliary qubit $a$, based on four pieces of information: (i) the quantum circuit, (ii) the auxiliary group $\mathcal{A}$, (iii) the interaction group $\mathcal{I}$, and (iv) the main qubits that have already been allocated, $\mathcal{Q}$. Although the quantum circuit does not appear explicitly as a parameter, it is accessed through function calls. The algorithm can be divided into two parts: the first attempts to allocate the auxiliary qubit based on its interactions (lines~\ref{line:aux:p1:begin}--\ref{line:aux:p1:end}), with the aim of minimizing the creation of new interactions. If this step fails to identify a suitable candidate, the algorithm iterates over the remaining qubits to find a location for the auxiliary qubit. Since the decomposition was chosen based on the number of auxiliary qubits required, there will always be a main qubit available for allocation.

\begin{figure}[htpb]
  \centering
  \begin{minipage}{.85\linewidth}
    \begin{algorithm}[H]
      \caption{Select the main qubit where the auxiliary will be allocated.}
      \label{alg:aux}

      \begin{algorithmic}[1]
        \Function{SelectMain}{$a$, $\mathcal{A}$, $\mathcal{I}$, $\mathcal{Q}$}
        \ForAll {$i \in $ \Call{InteractionQubits}{$a$}} \label{line:aux:p1:begin}
        \ForAll {$c \in $ reversed \Call{InteractionQubits}{$i$}} \label{line:aux:for_c}

        \If {$c \in \mathcal{A} \cup \mathcal{Q}$ } \label{line:aux:not_in_aq}
        \State \textbf{continue}
        \EndIf

        \If {$\mathcal{I} \neq \emptyset$ \textit{AND} $c \notin \mathcal{I}$ \textit{OR} $c \in $ \Call{CleanQubits}{}} \label{line:aux:if}
        \State \textbf{return} $c$
        \EndIf

        \EndFor
        \EndFor \label{line:aux:p1:end}

        \ForAll {$c \in $ \Call{CleanQubits}{}} \label{line:aux:p2:begin}
        \If { $c \notin \mathcal{A}\cup \mathcal{Q}$}
        \State \textbf{return} $c$
        \EndIf
        \EndFor

        \ForAll {$c \in $ \Call{AllQubits}{}} \label{line:aux:forall}
        \If {$c \notin \mathcal{I} \cup \mathcal{Q}$} \label{line:aux:forall:if}
        \State \textbf{return} $c$
        \EndIf
        \EndFor \label{line:aux:p2:end}

        \EndFunction

      \end{algorithmic}
    \end{algorithm}
  \end{minipage}
\end{figure}

To allocate an auxiliary qubit based on its interactions, Algorithm~\ref{alg:aux} first iterates over the qubits $i$ that interact with $a$ (line~\ref{line:aux:p1:begin}). The function \textsc{InteractionQubits} returns a list of qubits that interact with a given qubit via two-qubit gates. Note that \textsc{InteractionQubits}($a$) $\subseteq \mathcal{A} \cup \mathcal{I}$, so these qubits are not candidates to host the auxiliary qubit $a$, but they can guide the search for suitable candidates. The algorithm then iterates over the qubits in \textsc{InteractionQubits}($i$) in reverse order (line~\ref{line:aux:for_c}). The intuition is that a qubit $c$ that interacts with the same qubits as $a$ may be a good candidate to host $a$.

The algorithm first examines the qubits in \textsc{InteractionQubits}($a$) in the order of their interactions, seeking to identify the main qubit that interacts first with $a$. Next, it reviews the qubits in \textsc{InteractionQubits}($i$) in reverse order to find a main qubit whose final interaction aligns with the initial interaction of $a$. If the candidate qubit $c \notin \mathcal{A} \cup \mathcal{Q}$ (line~\ref{line:aux:not_in_aq}), it is suitable to host the auxiliary qubit $a$, provided that, if $a$ is a dirty auxiliary qubit ($\mathcal{I} \neq \emptyset$), $c$ is not in the interaction group (line~\ref{line:aux:if}). Alternatively, if $a$ is a clean auxiliary qubit, it can be allocated if $c$ is clean (line~\ref{line:aux:if}). It is important to note that a clean qubit used as an auxiliary will not be considered dirty afterward. In line~\ref{line:aux:p1:begin}, if $i \in \mathcal{A}$, then $\forall c \in \textsc{InteractionQubits}(i)$, $c \in \mathcal{A} \cup \mathcal{I}$, so that this iteration can be skipped.

Note that the first part of the algorithm ensures that no new interactions between qubits are created. Specifically, if two qubits have never interacted before, the allocation will not occur. This can be easily verified: if $a$ interacts with $i$, which in turn interacts with $c$, allocating $a$ to $c$ will not introduce a new interaction. However, this first step does not always lead to an allocation. For example, if the first multi-qubit gate is decomposed using clean auxiliary qubits, the algorithm may not find an allocation for $a$ during this initial phase. This can be easily checked: if the gate being decomposed is the first multi-qubit gate of the circuit, then for $c \in \textsc{InteractionQubits} (i)$, $c \in \mathcal{A}$ or $c$ is a dirty qubit, since every interaction is a result of the decomposition.

If no suitable candidate for the auxiliary qubit allocation is found, the algorithm guarantees a successful allocation in the next part (lines~\ref{line:aux:p2:begin}--\ref{line:aux:p2:end}). First, it iterates over qubits that have not been allocated or modified by the programmer--the clean qubits (line~\ref{line:aux:p2:begin}). If the decomposition selection phase (Subsection~\ref{subsec:selct}) specified the use of clean qubits, an allocation spot for the auxiliary qubit is guaranteed. If this attempt fails, the algorithm iterates over all available main qubits (line~\ref{line:aux:forall}), selecting the first one that is not in the interaction group or has not yet been allocated for this auxiliary group (line~\ref{line:aux:forall:if}).

In the next section, we will evaluate the performance of the automated selection of auxiliary qubits, focusing on its impact on reducing the number of CNOT gates. We will compare the implementation of quantum algorithms using Ket~\cite{darosaKetQuantumProgramming2022} with an equivalent implementation using Qiskit~\cite{javadi-abhariQuantumComputingQiskit2024}. Additionally, we will highlight how Ket's high-level constructs can enhance code readability and reduce the number of lines of code required.

\section{Performance Evaluation}
\label{sec:perf}

In this section, we evaluate the performance of our proposal by measuring the number of CNOT gates in the final quantum circuit after decomposition. The effectiveness of our approach is closely tied to the specific quantum algorithm being implemented, particularly when the algorithm requires the decomposition of multi-controlled single-target gates. For instance, variational quantum algorithms~\cite{blekosReviewQuantumApproximate2024} may not be significantly impacted by our proposal, as they primarily use single-qubit and two-qubit gates, typically CNOTs and $R_{ZZ}$ rotations, where the latter involves $R_Z$ rotations around CNOT gates.

Algorithms where automatic auxiliary qubit allocation significantly impacts the final results include, but are not limited to, Grover's quantum search algorithm~\cite{groverFastQuantumMechanical1996}, arbitrary quantum state preparation~\cite{kitaevWavefunctionPreparationResampling2008}, algorithms for solving systems of linear equations~\cite{harrowQuantumAlgorithmLinear2009}, and algorithms for solving differential equations~\cite{huQuantumCircuitsPartial2024}. In this section, we focus on evaluating the performance of our approach using two algorithms: Grover's algorithm (Subsection~\ref{subsec:grover}) and the state preparation algorithm depicted in Figure~\ref{subfig:prepare} (Section~\ref{subsec:prepare}).

As a basis for comparison, we use IBM's Qiskit platform~\cite{javadi-abhariQuantumComputingQiskit2024}, one of the most widely used quantum programming platforms, with over 500 contributors on GitHub. Qiskit provides a Python API that enables the construction of quantum circuits and the development of quantum applications. For our tests, we used Qiskit version 1.2.0.

While both Ket and Qiskit offer toolboxes for developing quantum applications, the strategies each platform employs for expressing quantum instructions differ. In Qiskit, the primary quantum programming structure is the quantum circuit, where the programmer must explicitly append quantum gates. In contrast, Ket's primary quantum structure is the \emph{Quant}, which represents a list of qubit references that can be passed to quantum gates. Along with the analysis of decomposition performance, we will highlight these differences and discuss their impact on high-level quantum programming.

Although circuit depth is the primary metric for determining execution time on a quantum computer, it is closely tied to the number of CNOT gates. The number of CNOTs directly influences factors such as the required SWAP operations during the quantum circuit mapping, the CPU time needed for circuit generation and compilation, and crosstalk errors in quantum computers~\cite{sarovarDetectingCrosstalkErrors2020}. Given these considerations, we have chosen to use the total number of CNOT gates as the primary metric for our comparisons.

\subsection{Grover's Algorithm}
\label{subsec:grover}
Grover's Algorithm is a seminal work in quantum computing, as it was one of the first to demonstrate that a quantum computer can outperform a classical computer in specific tasks. The algorithm is particularly effective in searching for a ``needle in a haystack'', allowing the identification of an entry in an unordered list of $N$ elements in time $O(\sqrt{N})$. Although Grover's Algorithm does not offer a superpolynomial speedup like Shor's factorization algorithm, it has significant implications in fields such as optimization~\cite{durrQuantumAlgorithmFinding1996}, cryptography~\cite{aaronsonQuantumLowerBounds2004,brassardQuantumCryptanalysisHash1998}, and solving NP-complete problems~\cite{cerfNestedQuantumSearch2000}.

The algorithm operates in a loop, where each iteration requires at least one multi-controlled Z gate to implement the Grover diffusion operator, and usually another multi-controlled gate for the oracle. The oracle is the part of the algorithm that encodes the logic of the problem being solved. In our evaluations, we implemented a simple oracle that marks the state $\ket{11\cdots1}$ using a multi-controlled Z gate.

Figure~\ref{fig:grover} illustrates the implementation of Grover's Algorithm in Ket (Figure~\ref{subfig:grover:ket}) and Qiskit (Figure~\ref{subfig:grover:qiskit}). In Ket, the algorithm takes a list of qubits as a parameter and returns a measurement, with all operations performed directly on the qubits. In Qiskit, the function takes the quantum circuit and the number of qubits as parameters, appending operations to the circuit. Consequently, Qiskit's function implementation does not return a value, as the operations are embedded in the circuit, including the measurement. The multi-controlled operation in Ket is implemented using the \texttt{ctrl} function, which can be applied to any callable, while in Qiskit, a new quantum gate is created using the \texttt{control} method from the \texttt{ZGate} class. Additionally, Ket allows for reducing the number of lines of code through the \texttt{with around} statement. Before decomposition, both implementations result in the same quantum circuit, despite the Ket implementation requiring fewer calls to apply quantum gates.

\begin{figure}[htbp]
  \centering
  \begin{subfigure}[t]{0.495\textwidth}
    \centering
    \begin{minipage}[t]{.98\linewidth}
      \begin{minted}[frame=lines,fontsize=\scriptsize ,breaklines]{py}
def grover(qubits: Quant) -> Measurement:
    # Apply Hadamard gates to all qubits
    H(qubits)
    # Number of Grover iterations
    steps = int((pi / 4) * sqrt(2**len(qubits)))
    for _ in range(steps):
        # Apply the Oracle
        ctrl(qubits[:-1], Z)(qubits[-1])



        # Apply the Grover diffusion operator
        with around(cat(H, X), qubits):
            ctrl(qubits[:-1], Z)(qubits[-1])






    # Measurement
    return measure(qubits)



      \end{minted}
    \end{minipage}

    \caption{Ket implementation.}
    \label{subfig:grover:ket}
  \end{subfigure}
  \begin{subfigure}[t]{0.495\textwidth}
    \centering
    \begin{minipage}[t]{.98\linewidth}
      \begin{minted}[frame=lines,fontsize=\scriptsize,breaklines]{py}
def grover(qc: QuantumCircuit, num_qubits: int):
    # Apply Hadamard gates to all qubits
    qc.h(range(num_qubits))
    # Number of Grover iterations
    steps = int((pi / 4) * sqrt(2**num_qubits))
    for _ in range(steps):
        # Apply the Oracle
        qc.append(
            ZGate().control(num_qubits - 1),
            range(num_qubits),
        )
        # Apply the Grover diffusion operator
        qc.h(range(num_qubits))
        qc.x(range(num_qubits))
        qc.append(
            ZGate().control(num_qubits - 1),
            range(num_qubits),
        )
        qc.x(range(num_qubits))
        qc.h(range(num_qubits))
    # Measurement
    qc.measure(
        range(num_qubits),
        range(num_qubits),
    )
      \end{minted}
    \end{minipage}

    \caption{Qiskit implementation.}
    \label{subfig:grover:qiskit}
  \end{subfigure}

  \caption{Grover's Algorithm implementations in Ket and Qiskit.}
  \label{fig:grover}
\end{figure}

\begin{figure}[htbp]
  \centering
  \begin{subfigure}[t]{0.495\textwidth}
    \centering
    \includegraphics[width=\linewidth]{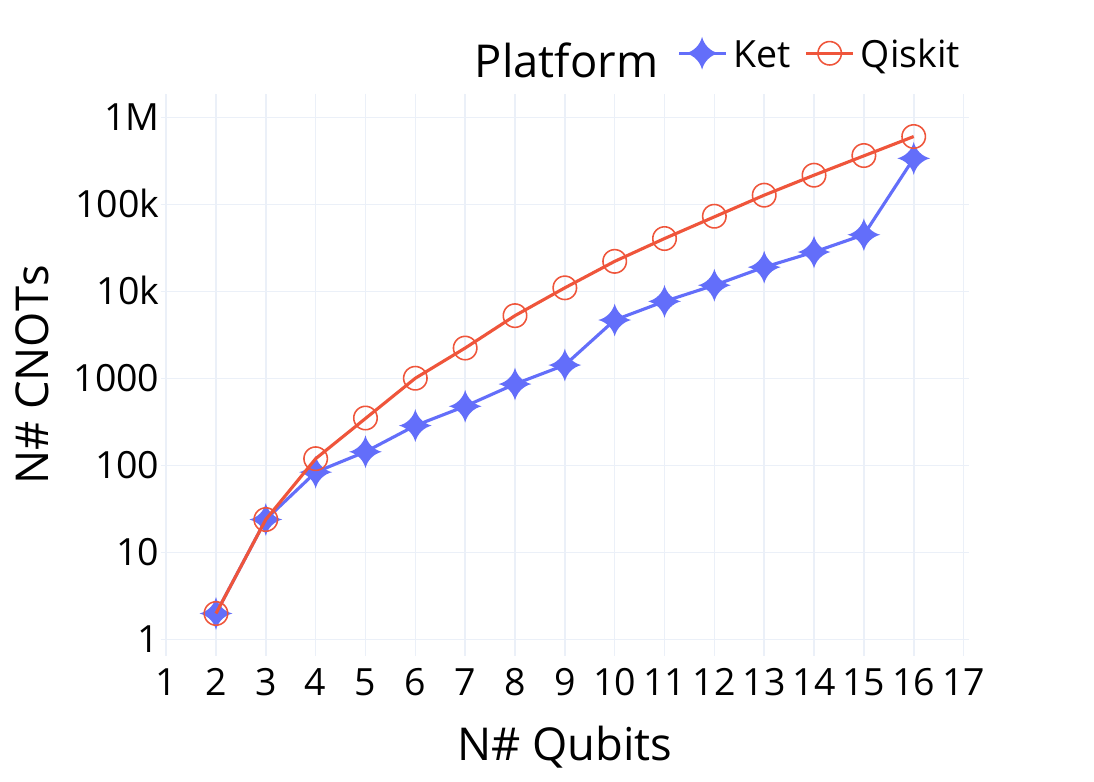}
    \caption{Number of CNOTs used.}
    \label{subfig:grover_time:sum}
  \end{subfigure}
  \begin{subfigure}[t]{0.495\textwidth}
    \centering
    \includegraphics[width=\linewidth]{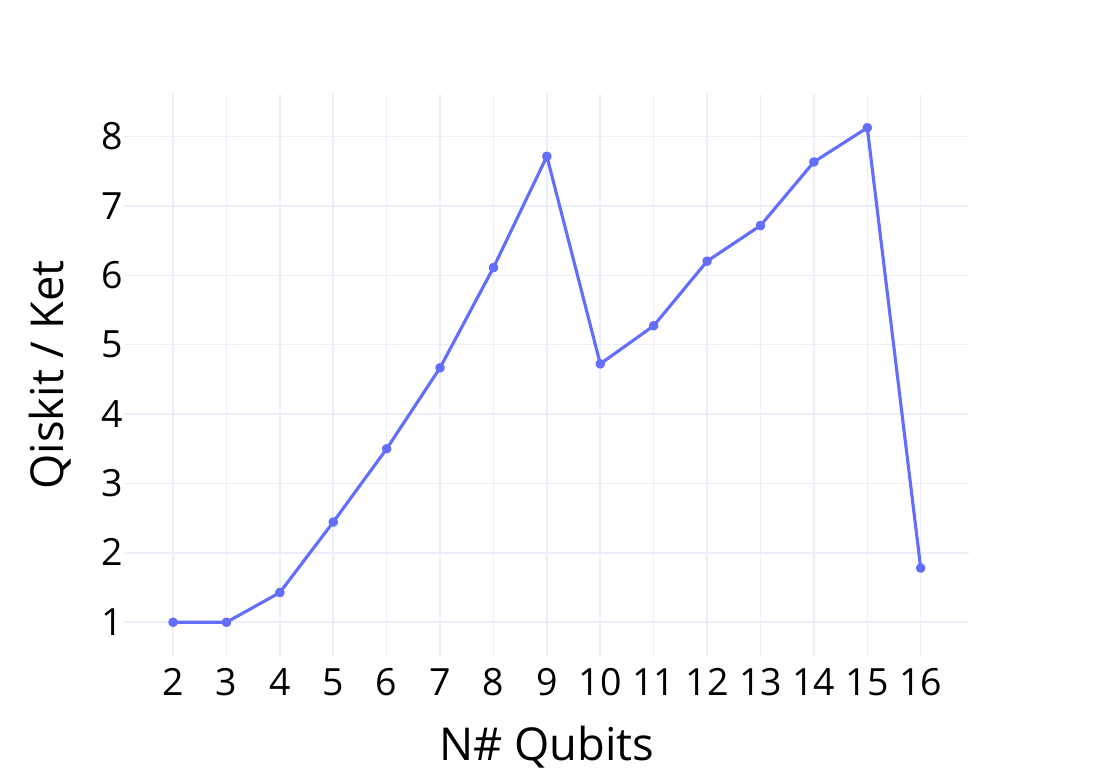}
    \caption{CNOT speedup (Ket vs. Qiskit).}
    \label{subfig:grover_time:speedup}
  \end{subfigure}

  \caption{Performance evaluation of Grover's Algorithm.}
  \label{fig:grover_time}
\end{figure}

For our experiment with Grover's Algorithm, we considered a quantum processing unit (QPU) with 16 qubits and executed the algorithm using between 2 and 16 qubits. Note that the implementations in Figure~\ref{fig:grover} can adapt to the number of qubits used by the algorithm and are independent of the QPU's total qubit count. Figure~\ref{fig:grover_time} presents the results of our experiment in terms of the number of CNOT gates used after decomposition. As shown in Figure~\ref{subfig:grover_time:sum}, Ket's compiled quantum circuit uses significantly fewer CNOT gates than the one resulting from Qiskit's execution, primarily due to Libket's ability to automatically allocate auxiliary qubits for decomposition.

In Figure~\ref{subfig:grover_time:speedup}, we present the speedup in terms of CNOT gates that Ket achieves over Qiskit in this experiment. The results can be divided into four phases, depending on the number of qubits used by the algorithm in relation to the QPU's available qubits. (i) First, for 2, 3, and 4 qubits, Ket offers no advantage over Qiskit, as the optimal decomposition without auxiliary qubits is employed. Note that for 4 qubits, there is a decomposition without auxiliary qubits that Qiskit seems to miss in this case, but this can easily be updated in future versions of Qiskit. (ii) From 5 to 9 qubits, the QPU has enough free auxiliary qubits available to perform the decomposition using the V-Chain circuit with clean auxiliary qubits. This can be observed in Figure~\ref{subfig:grover_time:speedup}, where Ket's speedup increases as more qubits are used, a result from the linear decomposition used by Ket in contrast with quadratic decomposition performed by Qiskit. (iii) With 9 qubits, Ket's performance significantly drops when the number of qubits needed for the V-Chain decomposition surpasses the total amount of auxiliary qubits available, forcing Libket to use the Pauli decomposition with a single clean auxiliary qubit. Still, Ket shows improvements up to 15 qubits. (iv) For 16 qubits, all qubits are being used by the algorithm, so no auxiliary qubits are available, resulting in a decomposition with a quadratic number of CNOTs. Note that Ket may still have an advantage over Qiskit, but only due to Ket implementing state-of-the-art quantum circuit decomposition, not due to any specific optimization implemented in Libket. Therefore, Qiskit could potentially update its decomposition algorithm for the \texttt{ZGate} without changing its architecture to match Ket's performance.

\begin{figure}[htbp]
  \centering
  \begin{subfigure}[t]{0.495\textwidth}
    \centering
    \includegraphics[width=\linewidth]{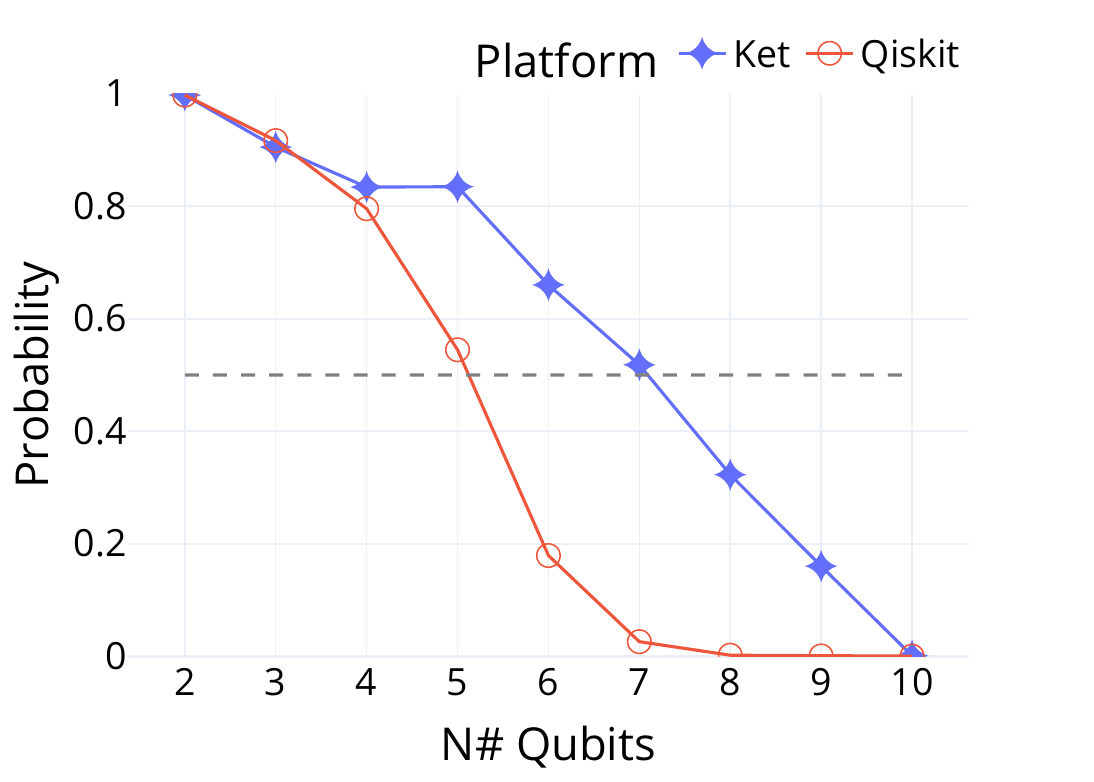}
    \caption{Probability of measuring the correct result.}
    \label{subfig:grover_time:noise}
  \end{subfigure}
  \begin{subfigure}[t]{0.495\textwidth}
    \centering
    \includegraphics[width=\linewidth]{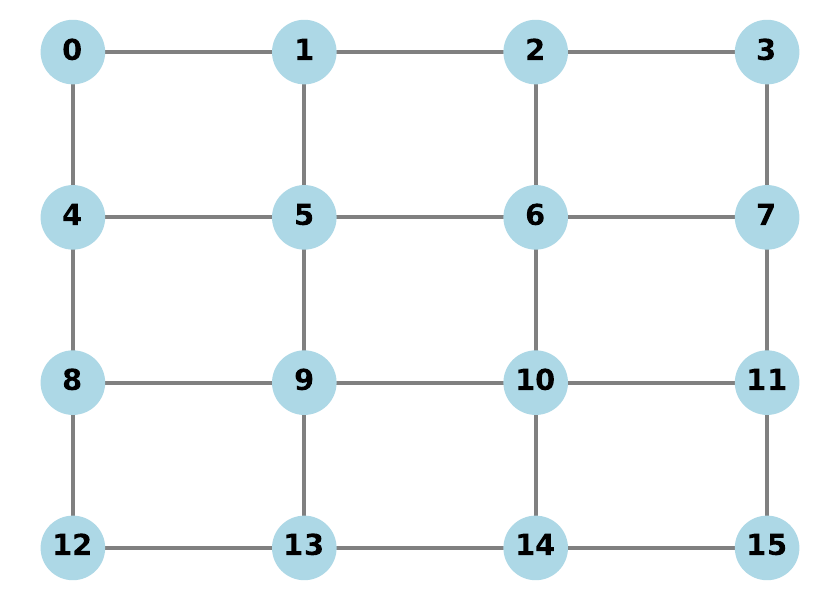}
    \caption{16-qubit QPU coupling graph.}
    \label{subfig:grover_time:grid}
  \end{subfigure}
  \caption{Noise simulation of Grover's Algorithm.}
  \label{fig:grover_noise}
\end{figure}

The number of CNOT gates has a direct impact on the performance of algorithms executed on noisy quantum computers~\cite{preskillQuantumComputingNISQ2018}. Since decoherence is closely tied to quantum circuit depth, reducing the number of CNOT gates can potentially enable the execution of quantum algorithms with larger inputs. We evaluated our proposal using a noisy quantum simulator to assess how the probability of measuring the correct result from Grover's Algorithm is affected by decoherence as the number of qubits increases. Figure~\ref{subfig:grover_time:noise} presents the results, comparing our proposal implemented in Ket with the performance of Qiskit. We observe that with just 6 qubits, the probability of measuring the correct state in Qiskit drops below 50\%, whereas in Ket, the probability remains above 50\% until 8 qubits are used.

For evaluating the effect of noise, we relied on the Qiskit Aer noise simulator for both Ket's and Qiskit's executions. In both cases, we considered a 16-qubit QPU with the coupling graph illustrated in Figure~\ref{subfig:grover_time:grid}, featuring the basis gates CNOT, $\sqrt{X}$, and $R_Z$. It is important to note that the strategies employed in Ket and Qiskit for mapping quantum interactions to the basis gates, and subsequently onto the coupling graph, may differ. In our simulations, we considered a CNOT error rate and duration of $7.992\times 10^{-8}$ and $10^{-5}$s, respectively, and an $\sqrt{X}$ error rate and duration of $2.997\times10^{-8}$ and $9\times10^{-5}$s. The $R_Z$ gate was assumed to have no error or duration, since it can be implemented virtually~\cite{mckayEfficientGatesQuantum2017}. Readout error was not considered.

\subsection{State Preparation}
\label{subsec:prepare}

The state preparation algorithm allows the creation of any quantum state $\sum_{k} \sqrt{r_k} \cdot e^{i\theta_k}\ket{k}$, given the list of measurement probabilities $[r_k]$ and the corresponding list of phases $[\theta_k]$ for the amplitude probabilities. This algorithm does not necessarily solve a computational problem but enables the preparation of an arbitrary quantum state, which is useful for performing experiments with quantum computers and testing other quantum algorithms with specific initial states~\cite{zanettiSimulatingNoisyQuantum2023}. The time complexity of this algorithm grows exponentially with the number of qubits, as the lengths of the lists $[r_k]$ and $[\theta_k]$ also increase exponentially. Since the state preparation algorithm heavily relies on the decomposition of multi-controlled gates, it requires a significant number of CNOT gates for execution. Optimizing the execution of this algorithm is essential to make it feasible on quantum hardware.

Figure~\ref{subfig:prepare} presents the implementation of the state preparation algorithm in Ket. It is worth noting that while there is no explicit call for a multi-controlled quantum gate, the recursive call to the \texttt{prepare} function within a controlled scope results in multi-controlled $R_Y$ and Phase gates. Note that the gates inside the \texttt{with around} statement do not generate controlled gates due to Ket's optimization~\cite{rosaOptimizingGateDecomposition2024}. The function \texttt{prepare} in Figure~\ref{subfig:prepare} is treated as a quantum gate in Ket. 

We implemented the state preparation algorithm in Qiskit using a similar strategy to Ket, as shown in Figure~\ref{subfig:prepara:qiskit}. Unlike the implementation in Ket, where the function acts as a quantum gate, the Qiskit implementation returns a quantum gate as an instance of the \texttt{Gate} class. The method \texttt{to\_gate} from the \texttt{QuantumCircuit} class is used to convert a quantum circuit into a quantum gate. Since the gate is constructed recursively, we can see in the \texttt{prepare} function itself how to append the gate to the quantum circuit.

\begin{figure}[htbp]
  \centering
  \begin{subfigure}[t]{0.495\textwidth}
    \centering
    \begin{minipage}[t]{.98\linewidth}
      \begin{minted}[frame=lines,fontsize=\scriptsize,breaklines]{py}
class ParamTree:
    def __init__(
            self,
            prob: list[float],
            amp: list[float]
        ):
        total = sum(prob)
        prob = [p / total for p in prob]
        l_prob = prob[: len(prob) // 2]
        l_amp = amp[: len(prob) // 2]
        r_prob = prob[len(prob) // 2 :]
        r_amp = amp[len(prob) // 2 :]

        self.value = sum(r_prob)
        self.value = 2 * asin(sqrt(self.value))
        if len(prob) > 2:
            self.left = ParamTree(l_prob, l_amp)
            self.right = ParamTree(
                r_prob,
                r_amp,
            )
        else:
            self.left = None
            self.right = None
            self.phase0 = amp[0]
            self.phase1 = amp[1]

    def is_leaf(self):
        return (self.left is None
            and self.right is None)
      \end{minted}
    \end{minipage}
    \caption{Class used in both Ket and Qiskit implementation.}
    \label{subfig:prepara:ket}
  \end{subfigure}
  \begin{subfigure}[t]{0.495\textwidth}
    \centering
    \begin{minipage}[t]{.98\linewidth}
      \begin{minted}[frame=lines,fontsize=\scriptsize,breaklines]{py}
def prepare(
    num_qubits: int,
    prob: ParamTree | list[float | int],
    amp: list[float] | None = None,
) -> Gate:
    if not isinstance(prob, ParamTree):
        prob = ParamTree(prob, amp)
    qc = QuantumCircuit(num_qubits)
    qc.ry(prob.value, 0)
    if prob.is_leaf():
        qc.x(0)
        qc.p(prob.phase0, 0)
        qc.x(0)
        qc.p(prob.phase1, 0)
        return qc.to_gate()
    qc.append(
        prepare(
            num_qubits - 1,
            prob.left,
        ).control(ctrl_state=0),
        range(num_qubits),
    )
    qc.append(
        prepare(
            num_qubits - 1,
            prob.right,
        ).control(),
        range(num_qubits),
    )
    return qc.to_gate()
      \end{minted}
    \end{minipage}
    \caption{Qiskit implementation.}
    \label{subfig:prepara:qiskit}
  \end{subfigure}

  \caption{State preparation algorithm in Qiskit. For the Ket implementation, see Figure~\ref{subfig:prepare}.}
  \label{fig:prepare_alg}
\end{figure}

As the \texttt{prepare} function from Figure~\ref{subfig:prepara:qiskit} returns a \texttt{Gate} instance, we can use the \texttt{control} method to append this gate in a controlled manner to the quantum circuit, similar to what is done with \texttt{ZGate} in Figure~\ref{subfig:grover:qiskit}. The semantics of appending the controlled version of the gate are equivalent to the recursive call within the \texttt{with control} block in Ket. In Qiskit, we can change the control state of a gate using the \texttt{ctrl\_state} argument of the \texttt{control} method, as seen in the first recursive gate call. In Ket, the control state is always $\ket{1}$, but we can achieve control over the $\ket{0}$ state using the instruction \texttt{with around(X, <control\_qubits>)}, as shown in the first recursive call in Figure~\ref{subfig:prepare}, resulting in a quantum circuit equivalent to the one generated by Qiskit.

Although the codes in Figure \ref{subfig:prepare} and \ref{subfig:prepara:qiskit} result in the same number of quantum gate calls--considering that gates inside the \texttt{with around} block are called twice--the final quantum circuits before decomposition are not identical. This discrepancy is due to the high-level quantum programming construct \texttt{with around}~\cite{rosaOptimizingGateDecomposition2024}, which allows Libket to optimize out the control qubits in the Pauli~$X$ gates of the recursive controlled calls. The Pauli~$X$ gates are used in the base case of the recursion (\texttt{is\_leaf}) and to flip qubits during controlled calls when the control state is $\ket{0}$. This highlights how high-level quantum programming constructs can simplify quantum application development and result in more efficient programs.

\begin{figure}[htbp]
  \centering
  \includegraphics[width=.99\linewidth]{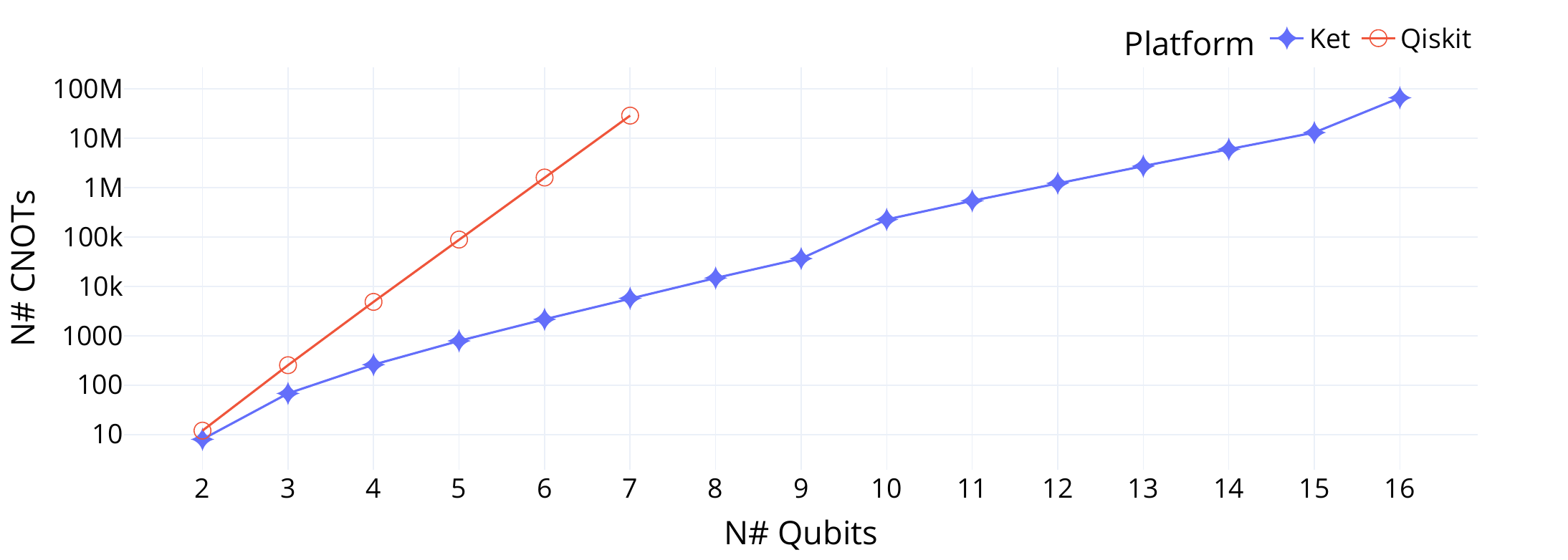}
  \caption{Performance evaluation of State Preparation Algorithm.}
  \label{subfig:prepara:sum}
  \label{fig:prepara}
\end{figure}

We evaluated the performance of the state preparation algorithm by measuring the number of CNOT gates in the final quantum circuit, using a QPU with 16 qubits. Figure~\ref{subfig:prepara:sum} shows how the number of CNOT gates increases as the number of qubits grows. The execution involves decomposing multi-controlled $R_Y$ and Phase gates, and in Qiskit, it also requires decomposing multi-controlled Pauli~$X$ gates. Ket demonstrates a significant improvement over Qiskit, primarily due to the automatic allocation of auxiliary qubits implemented in Libket. As shown in Figure~\ref{subfig:prepara:sum}, the Qiskit implementation exceeds 10 million CNOT gates with 7 qubits, whereas the Ket implementation only reaches this level with 15 qubits. It is worth noting that execution with 15 qubits processes 256 times the number of parameters compared to execution with 7 qubits. We limited the Qiskit execution to 7 qubits due to platform issues beyond this point, though we believe this is sufficient to illustrate the performance difference between the platforms.

Unlike Grover's algorithm, discussed in Subsection~\ref{subsec:grover}, which uses a single decomposition algorithm at a time, the state preparation algorithm requires decomposing various classes of controlled gates, including single-controlled gates, multi-controlled rotation gates, and multi-controlled Phase gates. Table~\ref{tab:alg_cnot} present the impact of different decomposition algorithms on the total number of CNOT gates. Note that as the number of qubits increases, five different decomposition algorithms are used by Libket.

\begin{table}[htbp]
  \centering
  \caption{Number of CNOT gates contributed by each decomposition algorithm during the execution of the state preparation algorithm in Ket on a 16-qubit QPU.}
  \label{tab:alg_cnot}
  \small
  \begin{tabular}{crrrrrr}
    \toprule
    &  \multicolumn{5}{c}{Decomposition Algorithm}\\
    N\# Qubits & CU(2) & Network & \multicolumn{1}{c}{SU(2)} & SU(2) Rewrite & Linear Depth & \multicolumn{1}{c}{Total} \\
    \cmidrule(rl){2-6}
    2   & \texttt{8} & -- & -- & -- & -- & \texttt{8} \\
    3   & \texttt{4} &   \texttt{64} & -- & -- & -- & \texttt{68} \\
    8   & \texttt{4} & \texttt{14,782} & -- & -- & -- & \texttt{14,786} \\
    9   & \texttt{4} & \texttt{36,676} & -- & -- & -- & \texttt{36,680} \\
    10  & \texttt{4} &  \texttt{8,208} &  \texttt{88,064} & \texttt{130,696} & -- & \texttt{226,972} \\
    15  & \texttt{4} &    \texttt{32} & \texttt{6,028,784} & \texttt{7,053,480} & -- & \texttt{13,082,300} \\
    16  & \texttt{4} & --              & \texttt{13,106,688} & -- & \texttt{53,184,336} & \texttt{66,291,028} \\
    \bottomrule
  \end{tabular}
\end{table}

We can divide the growth in the number of CNOT gates as the number of qubits used--out of the 16 available--increases into four phases: (i) First, with only two qubits, only the single-control $U(2)$ gate decomposition is required. As shown in Table~\ref{tab:alg_cnot}, the single-control $U(2)$ gate decomposition is applied in every execution. (ii) Between 3 and 9 qubits, any quantum gate involving more than two qubits can be decomposed using the Network decomposition, as there are enough clean qubits available in the QPU to be used as auxiliary qubits. (iii) Between 10 and 15 qubits, some multi-controlled gate decompositions cannot be performed using the Network algorithm. Therefore, certain multi-controlled $R_Y$ gates must be decomposed using the $U(2)$ decomposition, which does not require auxiliary qubits. Additionally, multi-controlled Phase gates must be decomposed using the $SU(2)$ Rewrite decomposition, which requires a single clean auxiliary qubit. It is worth noting that the $SU(2)$ Rewrite decomposition accounts for more than half of the CNOT gates in the final quantum circuit. (iv) With 16 qubits, there are no free auxiliary qubits for decomposing multi-controlled Phase gates using the $SU(2)$ Rewrite algorithm. Consequently, Libket resorts to the Linear Depth decomposition, which is responsible for more than 80\% of the CNOT gates in the final quantum circuit.

\section{Conclusion}
\label{sec:conclusion}

In this paper, we demonstrated how the abstraction provided by high-level quantum programming can enhance the performance of quantum applications. Specifically, we showed that any multi-qubit gate can be decomposed into a finite set of single-qubit gates, provided that the programming platform supports their controlled versions. This approach, implemented by Libket—the runtime library of the Ket high-level quantum programming platform—enables the construction of complex quantum instructions, such as the Quantum Fourier Transform, and allows these instructions to be executed in a controlled manner. After establishing that the compiler requires only a limited set of decomposition algorithms to handle any multi-qubit gate, we demonstrated how the compiler can dynamically select the most efficient decomposition algorithm based on the current state of the quantum circuit and the target QPU. This capability results in more efficient quantum executions.

Our primary contribution is an algorithm that autonomously selects the optimal decomposing for a multi-qubit gates based on the availability of auxiliary qubits in the QPU. By leveraging unused qubits, this approach reduces the number of CNOT gates in the compiled circuit, thereby lowering the overall circuit depth. Additionally, our strategy for auxiliary qubit allocation minimizes the introduction of new interactions between qubits, positively impacting the subsequent compilation phase that maps the circuit onto the QPU's coupling graph.

We evaluated our proposal, implemented in the Ket platform, against Qiskit, one of the most widely used quantum programming platforms. Although Qiskit allows for appending controlled quantum gates in the quantum circuit, it does not make use of auxiliary qubits during decomposition, leading to decompositions that require a quadratic number of CNOT gates. Also, while Qiskit provides mechanisms to use auxiliary qubits to reduce circuit depth, the responsibility for selecting these qubits falls on the programmer, potentially resulting in suboptimal choices. Compared to Qiskit, Ket offers up to a quadratic reduction in the number of CNOT gates for multi-controlled gate decompositions, all without requiring any special programming instructions--completely transparent to the programmer.

Our proposal is not limited to the decomposition algorithms discussed in Section~\ref{sec:decom}. While we believe that the current implementation in Libket represents the state-of-the-art in quantum gate decomposition, future improvements can easily be incorporated as new algorithms are developed. These updates can be seamlessly integrated into Libket without requiring any architectural changes to the runtime library, further enhancing performance.

One of the primary objectives of our auxiliary qubit allocation algorithm is to minimize the creation of new qubit interactions, which, we believe, will lead to fewer SWAP operations during the circuit mapping compiling phase. However, the impact of our approach on the overall compilation process has not yet been thoroughly evaluated. As part of future work, we intend to compare the performance of our allocation algorithm against random selection.

Lastly, while this work emphasizes the use of auxiliary qubits for decomposing multi-qubit gates, other quantum algorithms may also benefit from the use of auxiliary qubits. A promising direction for future research is to allow programmers to explicitly allocate auxiliary qubits, while the compiler handles their management automatically, providing more flexibility and efficiency in quantum algorithm development.

\section*{Acknowledgement}

EID acknowledges the Conselho Nacional de Desenvolvimento Científico e Tecnológico - CNPq through grant number 409673/2022-6 and ECRR acknowledges the Coordenação de Aperfeiçoamento de Pessoal de Nível Superior - CAPES, Finance Code 001.

\bibliographystyle{unsrtnat}
\bibliography{main}

\end{document}